\documentclass[usenatbib,useAMS]{mn2e}
\usepackage{epsfig}
\usepackage{times}
\usepackage{amsmath}

\usepackage{verbatim}
\usepackage{color}

\def\gtrsim{\mathrel{\hbox{\rlap{\hbox{\lower3pt\hbox{$\sim$}}}\hbox{\raise2pt\hbox{$>$}}}}}
\def\lesssim{\mathrel{\hbox{\rlap{\hbox{\lower4pt\hbox{$\sim$}}}\hbox{$<$}}}} 
\newcommand{\simgt}%
        {\,\hbox{\lower0.6ex\hbox{$\sim$}\llap{\raise0.6ex\hbox{$>$}}}\,}
\newcommand{\simlt}%
        {\,\hbox{\lower0.6ex\hbox{$\sim$}\llap{\raise0.6ex\hbox{$<$}}}\,}

\newcommand{\aap}{A{\&}A}
\newcommand{\mnras}{MNRAS} 
\newcommand{\apj}{ApJ}
\newcommand{\apjl}{ApJ Lett.}
\newcommand{\apjs}{ApJ Supp.}
\newcommand{\nat}{Nature}
\newcommand{\aj}{AJ}
\newcommand{\pasp}{PASP}
\newcommand{\nar}{New Astron. Rev.}

\title[Brightness of violent WD mergers]{On the brightness distribution of Type Ia supernovae
  from violent white dwarf mergers}
\author[A.~J.~Ruiter et al.]{A. J. Ruiter$^{1}$\thanks{E-mail: \texttt{ajr@mpa-garching.mpg.de}}, S. A. Sim$^{2}$, R. Pakmor$^{3}$,
  M. Kromer$^{1}$, I. R. Seitenzahl$^{1,4}$, K. Belczynski$^{5,6}$,
\newauthor M. Fink$^{1,4}$, M. Herzog$^{1}$, W. Hillebrandt$^{1}$, 
F. K. R\"{o}pke$^{1,4}$, S. Taubenberger$^{1}$\\
$^{1}$Max-Planck-Institut f\"{u}r Astrophysik, Karl-Schwarzschild-
Str. 1, D-85741 Garching, Germany\\
$^{2}$Research School of Astronomy and Astrophysics, The Australian
National University, Mount Stromlo Observatory, Cotter
Road, Weston Creek, ACT 2611,\\ Australia\\
$^{3}$Heidelberger Institut f\"{u}r Theoretische Studien, Schloss-
Wolfsbrunnenweg 35, 69118 Heidelberg, Germany\\
$^{4}$Institut f\"{u}r Theoretische Physik und Astrophysik, Universit\"{a}t W\"{u}rzburg, Am Hubland, D-97074 W\"{u}rzburg, Germany\\
$^{5}$Astronomical Observatory, University of Warsaw, Al.
            Ujazdowskie 4, 00-478 Warsaw, Poland\\
$^{6}$Center for Gravitational Wave Astronomy, University of Texas at
            Brownsville, Brownsville, TX 78520, USA
}

\date{\today}

\pagerange{\pageref{firstpage}--\pageref{lastpage}}
\pubyear{}
\begin{document}
\maketitle
\label{firstpage}

\begin{abstract}
We investigate the brightness distribution expected for thermonuclear
explosions that might result from the ignition of a detonation
during the violent merger of 
white dwarf (WD) binaries.  Violent WD mergers are a subclass of the 
canonical double degenerate scenario where two carbon-oxygen 
(CO) WDs merge when the larger WD fills its Roche-lobe. 
Determining their brightness distribution is critical 
for evaluating whether such an explosion model could 
be responsible for a significant fraction
of the observed population of Type~Ia supernovae (SNe Ia).
We argue that the brightness of an explosion realised via the violent merger
model is mainly determined by the mass of $^{56}$Ni produced in 
the detonation of the primary CO~WD. To quantify this link,
 we use a set of sub-Chandrasekhar mass WD detonation models to
derive a relationship between primary WD mass ($m_{\rm WD}$) and 
expected peak bolometric brightness ($M_{\rm bol}$).
We use this $m_{\rm WD}$-$M_{\rm bol}$ relationship to convert the
masses of merging primary WDs from binary population models to a
predicted distribution of explosion brightness. 
We also investigate the sensitivity of our results to
assumptions about the conditions required to realise a detonation during 
violent mergers of WDs.
We find a striking similarity between the shape of our 
theoretical peak-magnitude distribution and that observed for 
SNe Ia: our model produces a $M_{\rm bol}$
distribution that roughly covers the range and matches the shape of
the one observed for SNe Ia.
However, this agreement hinges on a particular phase 
of mass accretion during binary evolution:  
the primary WD gains ${\sim}0.15-0.35$ M$_{\odot}$ from a
slightly-evolved helium star companion.  
In our standard binary evolution model, 
such an accretion phase is predicted to occur for about $43$\%
of all binary systems that ultimately give rise to binary 
CO~WD mergers. We also find that with high
  probability, violent WD mergers involving
  the most massive primaries (${\gtrsim}1.3 M_{\odot}$, which should
 produce bright SNe) have delay times ${\lesssim}500$ Myr.
\end{abstract}


\section{Introduction}
\label{sec:intro}

It is widely agreed that the progenitors of Type Ia supernovae (SNe~Ia) 
are thermonuclear explosions of carbon-oxygen (CO) white dwarfs (WDs) in
binary systems. 
However, it is still not clear whether their companion stars (secondaries)
are hydrogen-rich (main sequence or giant stars, typically
called `single degenerate' scenarios), helium-rich (helium
stars or helium-rich WDs) or other CO~WDs which merge with the primaries
(typically called `double degenerate' scenarios).

\citet{Ste11} analysed high-resolution spectra of 35 SNe~Ia, 
finding that more than half of the spectra exhibit a blueshifted 
Na I D absorption feature.  The authors attribute this feature to 
circumstellar material (blown off of the secondary prior to the SN
explosion), suggesting that at least 20\% of SNe~Ia in spiral 
galaxies originate from single degenerate scenarios \citep[see
also][]{Pat11}.  
On the other hand, several recent studies have highlighted challenges to
single degenerate scenarios and have instead tended to favour double
degenerate models. For example, the lack of emission associated
with the supernova overrunning the companion star 
\citep[][]{kasen10,hayden10,ganeshalingam11,Bia11}, 
the absence of radio detections \citep[][]{hancock11,Cho12,Hor12} and 
the lack of unambiguous identifications of companion stars in supernova
remnants \citep[][]{ruizlapuente04,kerzendorf09}
argue against 
models in which the 
companion is a red giant star at explosion \citep[see
also][where a single degenerate scenario origin for SN 2011fe is
likely ruled out]{Blo12}. 
In addition, a number of recent studies find that the delay time
distribution (DTD) of SNe~Ia follows a power-law shape $t^{s}$, where 
$-1.6 \le s \le -1$
\citep{MSG10,Gra11,Bar12,San12}.  Such a distribution is expected if the 
progenitor
population of SNe~Ia is dominated by double degenerates.
These studies, among others, have fuelled speculation
that there must be at least two different progenitor scenarios 
leading to SNe~Ia.  
However, the question of which channel(s) dominate(s) the production of
`normal' SNe~Ia
\citep{Ben05,BDB09} and determine(s) the observed 
SN~Ia brightness distribution remains open.

In a series of recent papers \citep{pakmor10,pakmor11,pakmor12,roepke12}, 
a new explosion model for SNe~Ia produced by double degenerate (CO+CO~WDs) 
mergers
was investigated. 
In this violent merger model, a prompt detonation is triggered during
the merger itself, leading to a thermonuclear explosion of the system.
Although it is not yet established which parameters of the merging
system are most crucial for the formation of such a detonation
\citep[see e.g.,][]{Dan12}, 
synthetic spectra and light-curves computed for several realisations 
of the violent merger
scenario are a good match to maximum light observations 
of SNe~Ia \citep{pakmor10,pakmor12}. 
The level of agreement with observations is comparable to that found for
other scenarios \citep{roepke12}. 

In contrast to Chandrasekhar mass explosion models,
where the peak
brightness is driven by stochastic processes connected to
the formation of deflagration and detonation flames or other
properties of the exploding Chandrasekhar mass WD
\citep[e.g.][]{kasen09,blondin11,SCR11},
in the violent merger scenario there is a direct correlation between
fundamental parameters of the progenitor system and the SN luminosity.
Specifically, to first order, the violent merger model can be interpreted as a 
mechanism to ignite a detonation
of a sub-Chandrasekhar-mass WD (the primary CO~WD) as described in
\citet{Sim10} \citep[see also][]{Fin10}. Although the merging secondary WD is also consumed by 
the thermonuclear flame of the detonation, it does not produce any 
iron group elements when it is burned. Thus, the absolute brightness 
of a SN~Ia explosion depends on the mass of the primary CO~WD.

The brightness distribution of SNe~Ia can be observed directly, 
with a typical peak brightness 
of ${\simeq}-19$~mag \citep{Ric02,Ben05}
and a spread in brightness of ${\sim}1$~mag.  
Recently, \citet{li11} presented the brightness  
distribution of a volume-limited sample of SNe~Ia.
Any theoretical scenario explaining SNe~Ia that claims to account for a
large fraction of observed events must be able to reproduce the
observed brightness distribution. 
The goal of this paper is to quantify the brightness 
distribution of SNe~Ia from the violent merger scenario.

Our study requires three main inputs.
First, we need predictions of the distribution of properties of merging
CO+CO~WD pairs. These we obtain from previously published {\sc
  StarTrack} \citep{BKB02,Bel08} 
binary population synthesis models, which are described in Section~\ref{sec:BPS}.
Second, we must consider which characteristics of a merging system 
will be required for the explosion mechanism to work. This 
specification will be made
based on the mass ratio of the merging system, as described in
Section~\ref{sec:qc}.
Third, we need to 
estimate the brightness of an explosion based on the properties
of the merging WDs. This is discussed in Section~\ref{sec:bright}.
In Section~\ref{sec:results}, we present our results based on the standard 
model used for the double degenerate scenario (DDS) in \citet{Rui11} and
comment on the comparison with observations. 
We discuss important implications and uncertainties of our modelling
in Section~\ref{sec:discuss} before drawing conclusions in Section~\ref{sec:conclusions}.

\section{Binary population synthesis}
\label{sec:BPS}

\subsection{Mass exchange for StarTrack binaries}

We consider the population synthesis calculation results
from the standard model of \citet{Rui11}, whose data were obtained
with the {\sc StarTrack} population synthesis binary evolution code.  
In this subsection, we briefly explain how stellar mass loss/gain 
is handled over the course of binary evolution in {\sc StarTrack}.

{\em Unstable mass transfer.} 
Outside densely populated environments -- such as the cores of globular
clusters -- binary stars that form close double WDs  
must undergo at least one 
common envelope (CE) phase \citep{IL93}, in which the size of the binary orbit 
decreases (often significantly) upon expulsion of the mass-losing star's envelope.
Currently, CE ejection remains a poorly understood process in 
astrophysics. Observationally, it is difficult to study since it is
a short-lived evolutionary phase\footnote{Though the sample of post-CE binaries (PCEBs) is becoming
  larger \citep{Zor11}.}
and detailed theoretical modelling is extremely challenging given the large 
spatial and temporal scales that must be taken into account 
to properly approach the problem \citep{RT12}.  However, despite our collective 
ignorance, the CE phase cannot be ignored in binary evolution modelling,
and so a simplified parametrization of the CE phase is often employed
in binary population synthesis codes. 

In the standard model of \citet{Rui11}, the 
CE parametrization $\alpha_{\rm CE}\lambda= 1$ was used. 
In this CE parametrization, often coined the `energy formalism' 
\citep{Web84}, it is assumed that the binary can expel the envelope
without merging if the binding energy
of the mass-losing star's envelope is not greater than the
orbital energy of the binary with some efficiency factor $\alpha_{\rm CE}$;
$\Delta E_{\rm bin} \le \alpha_{\rm CE} \, \Delta E_{\rm orb}$
\citep[but see also][for an overview and recent work on the CE problem]{IC11,Iva11}. 
Since our calculations
are based on the results of \citet{Rui11}, we assume as they do that 
the transfer of energy from the orbit to the removal of the envelope 
during CE is fully efficient, and thus $\alpha_{\rm CE} = 1$.  
The binding energy is inversely proportional to the term $\lambda$,
which is a parametrization of the structure of the mass-losing star 
\citep[e.g.][]{deK90}.  
$\lambda$ is somewhat uncertain, but is often assumed to be on the 
order of unity in population synthesis calculations
for low- and intermediate-mass stars on the main
  sequence (MS) or the giant branch \citep[][]{DT00}. 
We take the CE 
parametrization with $\alpha_{\rm CE}=1$ and $\lambda = 1$ 
from \citet{Rui11} for the current study. 

{\em Stable mass transfer.}
For Roche-lobe overflow (RLOF) on to non-degenerate accretors, mass
transfer is assumed to be non-conservative.  
An efficiency parameter -- which characterizes the 
fraction of matter transferred to the companion -- 
is adopted \citep[termed $f_{\rm a}$ in Equation (33) of][]{Bel08}.  
In \citet{Rui11}, $f_{\rm a}=0.5$ was adopted for the standard 
calculations \citep[see also][where the same parameter is termed `$\beta$']{Men10}. 
Thus only $50$\% of the matter lost by the donor is gained by
the accretor (Eddington-limited) while the rest is lost from the binary system with
the specific angular momentum of the orbit (see Appendix A for more
discussion). 

For WD accretors,
the {\sc StarTrack} code takes into account 
a number of published results on mass accretion on to CO~WDs for
hydrogen-rich accretion \citep{PK95,Nom07}, and helium-rich 
accretion \citep[][see also \citealt{Taa80,GBW99}]{KH99,KH04}.
In this case, the fraction of mass gained
by the accretor is dependent upon the rate at which mass is transferred from the
donor and the mass of the accreting WD.  Mass that is lost from the
binary carries with it the specific angular momentum of the accretor.  
A more detailed description can be found in \citet[][section 5.7.2]{Bel08}.

\subsection{Primary mass distribution}

\begin{figure}
\epsfig{file=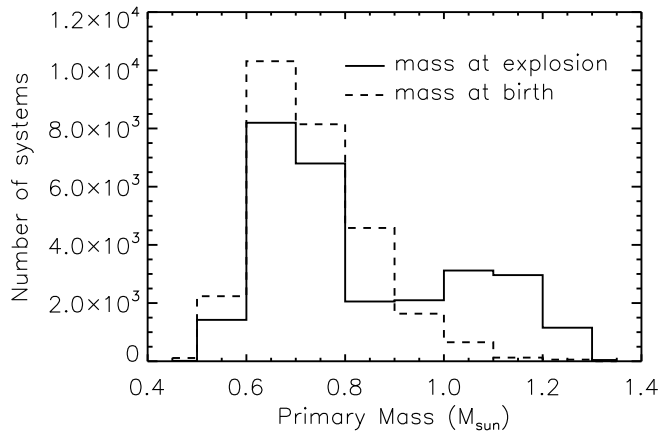,width=9cm}\\
\caption{The mass distribution of the primary WD for
all CO+CO~WD binaries at the time of the merger from the Ruiter et
  al. (2011) standard calculation (solid histogram). 
The dashed histogram shows the mass distribution of the primary when it
becomes a CO~WD (i.e. neglecting any mass accretion which may occur 
after the primary WD has formed, see text).
The numbers on
  the $y-$axis are not scaled, and show the actual numbers
  of CO+CO~WD mergers produced in a Hubble time in the simulation of 
  $4.5 \times 10^{6}$ zero-age main sequence binary stars.  
}
\label{fig:mass-dist-std}
\end{figure}

Fig.~\ref{fig:mass-dist-std} shows the distribution of primary
(more massive WD) masses for merging CO+CO~WD pairs at the time of merger 
from the \cite{Rui11} $\alpha_{\rm CE}\lambda = 1$ calculation
(solid histogram).
Note that this $m_{\rm WD}$ distribution
has two peaks: a prominent peak at ${\sim}0.65$
  M$_{\odot}$ and a smaller bump at ${\sim}1.1$ M$_{\odot}$.
For comparison, we show 
the distribution of primary masses for the same population
of merging CO+CO~WD pairs {\em recorded at the time of primary WD birth}
(dashed histogram); this distribution has only a single peak at
 ${\sim}0.65$ M$_{\odot}$.  This `mass shift' to larger masses is the 
result of a phase of mass accretion when the donor star is in 
the helium-burning phase, which we discuss in detail in 
Section~\ref{subsec:binev}.
As we will discuss in Section~\ref{sec:bright}, this has important 
implications for our predicted brightness distribution of 
SNe~Ia.

\subsection{Binary evolution and example}
\label{subsec:binev}

To understand the origin of
the double-peaked $m_{\rm WD}$-distribution, we have analysed various binary
evolution histories for representative systems in the {\sc StarTrack}
simulations. 
The formation of the secondary bump at ${\sim}1.1$ M$_{\odot}$ in the
solid histogram of Fig. 1 (and
more generally the larger average primary WD masses)
is associated with CO~WD primaries that accreted material from a 
slightly evolved naked helium star companion 
\citep[k=8 in][]{HPT00}. 

\begin{figure}
\epsfig{file=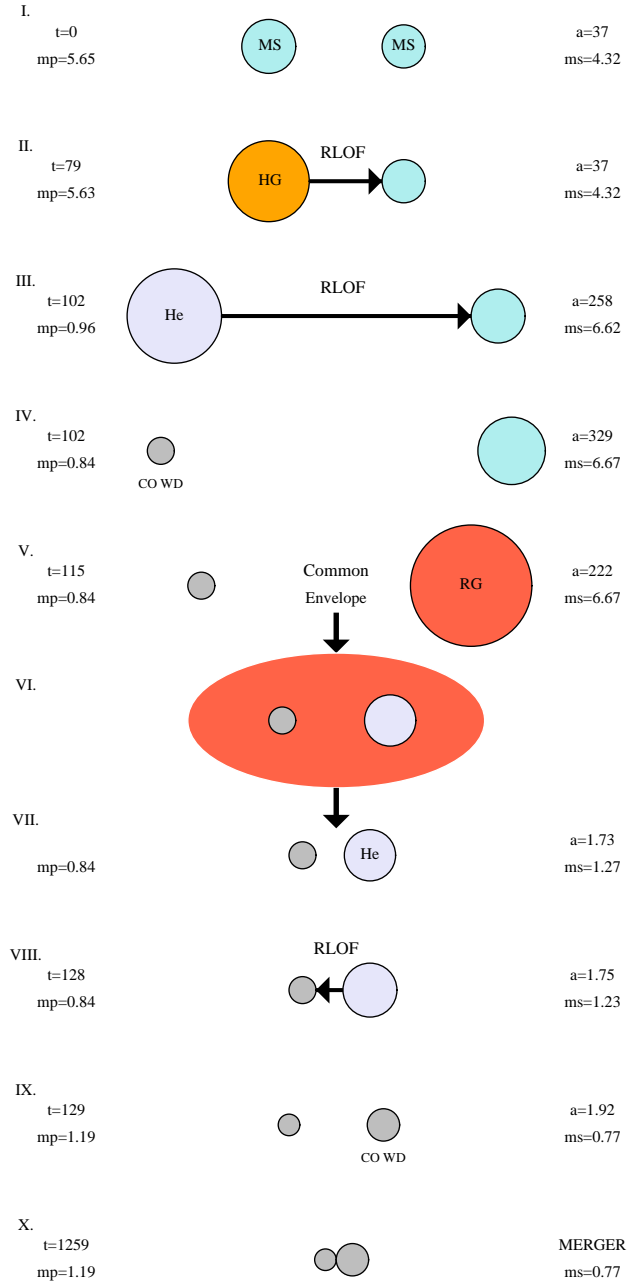,width=9cm}\\
\caption{Flowchart showing a typical evolutionary channel where the
  primary WD gains mass 
via RLOF from its helium star companion, prior to the double CO~WD
merger.  The duration of the RLOF phases and other evolutionary
details are given in the text.}
\label{fig:evol}
\end{figure}

\begin{figure}
\epsfig{file=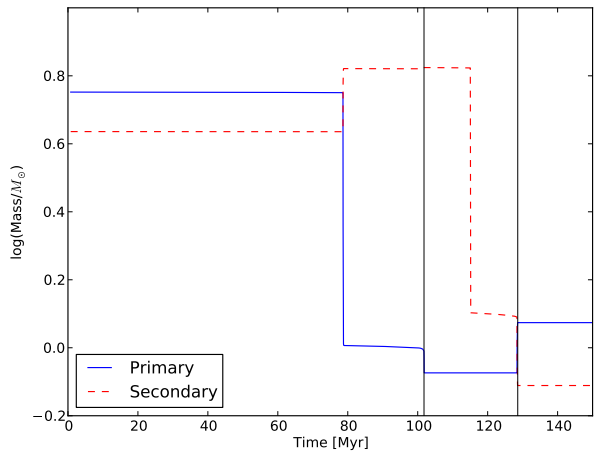,width=9cm}\\
\caption{Mass evolution for the binary system shown in
  Fig.~\ref{fig:evol}.  We do not show the last ${\sim}$Gyr of evolution
  since the WD masses remain unchanged during that time.  The common envelope event is marked by the
  drastic decrease in secondary mass at 115 Myr, while the two
  vertical lines indicate the formation times of the CO~WDs.}
\label{fig:masses}
\end{figure}

To be more specific, we present a typical evolutionary channel 
along with the relevant physical processes that may potentially 
lead to the formation of a 
CO+CO~WD merger that contributes to the bump  
at ${\sim}1.1$ M$_{\odot}$ in Fig.~\ref{fig:mass-dist-std}.  
In our example, the initially more massive star on the zero-age main
sequence (ZAMS) will become the first formed WD and the more massive WD 
at the time of the merger, though this is not always the case for 
all evolutionary channels. 
However, throughout the paper, we define the primary 
(with mass $m_{\rm p}$) as the star 
which ends up as the more massive 
WD at the time of the merger, and 
the companion is the secondary (with mass $m_{\rm s}$).  
We illustrate the onset of the most important evolutionary stages 
in Fig.~\ref{fig:evol} with Roman numerals (I--X), where time
$t$, masses $m_{\rm p}$ and $m_{\rm s}$, and separation $a$ are
indicated in Myr,
M$_{\odot}$ and R$_{\odot}$, respectively.  
We also show the mass evolution of the binary in 
Fig.~\ref{fig:masses}.

Our example begins with two MS stars on a circular orbit with 
an orbital period of $8.3$ h ($a$=37 R$_{\odot}$, {\bf stage I}). On the ZAMS,   
$m_{\rm p}$ and $m_{\rm s}$ are $5.65$ and $4.32$ M$_{\odot}$,
and their corresponding radii are $2.84$ and $2.43$ R$_{\odot}$, respectively.  
At $79$ Myr the primary evolves off the
MS and fills its Roche-lobe, leading to RLOF 
while in the Hertzsprung Gap ({\bf stage II}).\footnote{See
  \citet[][sect. 5.6]{Bel08} for a description of how stellar rejuvenation
  is treated in {\sc StarTrack}.} 
The mass ratio 
is fairly close to unity, so RLOF is stable ($f_{\rm a}=0.5$) though it
initially proceeds on a thermal time-scale.  
Over the course of RLOF, which lasts 0.15 Myr, there is a mass ratio reversal, where the
primary becomes the less massive star. 
The mass transfer (from less massive to more massive star) and
expansion of the primary 
drive an increase in separation to $233$ R$_{\odot}$, the primary  
shrinks back inside its Roche-lobe and RLOF ceases. 
By the time RLOF has stopped, $m_{\rm p}=1.02$ M$_{\odot}$ while 
$m_{\rm s}$ has increased to $6.62$ M$_{\odot}$.  
The primary continues to evolve up the red giant branch, though by this point
most of its hydrogen-rich layers have been removed in RLOF.
Once the hydrogen shell-burning phase is complete, the primary becomes a
naked helium star (core of red giant).   
At t=99 Myr, the primary becomes a slightly evolved naked helium star
and increases significantly in size.  At t=102 Myr, the primary
fills its Roche-lobe and a second 
phase of stable RLOF ensues ({\bf stage III}).
Helium-rich material is transferred to
the secondary on a nuclear time-scale for 0.14 Myr (as before 
$f_{\rm a}=0.5$). 
Mass transfer ceases at a separation of $a=318$ R$_{\odot}$, 
when $m_{\rm p}=0.86$ M$_{\odot}$ and 
$m_{\rm s}=6.67$ M$_{\odot}$.  By this stage in the evolution, 
the primary has converted much of its helium core into CO, and upon
completion of helium burning at $t=102$ Myr, it becomes a CO
WD ({\bf stage IV}).
   
At t=115 Myr, the primary begins its journey up the red giant branch.
Once the secondary reaches a radius of 123 R$_{\odot}$ (orbital period of 140 d),
a CE takes place ({\bf stages V, VI}), leaving behind 
the CO~WD and the (naked) core of the red giant, 
a decreased separation (orbital period
of 261 min) and $m_{\rm s}=1.27$ M$_{\odot}$ ({\bf stage VII}). 
At $t=127$ Myr, the secondary becomes a slightly evolved naked helium
star.   
Soon afterwards it fills its Roche-lobe which leads to a 
third RLOF phase ({\bf stage VIII}). 

The mass transfer rate initially proceeds on a thermal time-scale 
and is fairly high ($\dot{M}_{\rm tran}=8.8 \times 10^{-5}$ M$_{\odot}$ yr$^{-1}$).
For the given WD accretor mass, such a mass transfer rate
is above the critical threshold that leads to 100\% burning efficiency
\citep[the critical value is $4.57 \times 10^{-7}$ M$_{\odot}$ yr$^{-1}$
for $0.84 $ M$_{\odot}$ WD, interpolated from][Equation 6]{KH04}.  
However, the mass
transfer rate is super-Eddington, and so the accretion rate is capped 
at the Eddington accretion rate 
($\dot{M}_{\rm accr} = 2.0 \times 10^{-5}$M$_{\odot}$ yr$^{-1}$;
Eddington accretion lasting $0.002$
  Myr).\footnote{If such binaries 
    lead to SNe Ia, we estimate  
    ${\sim}$few to 10 systems to be in this
    high-accretion phase and thus bright and potentially visible in
    the Galaxy \citep[Galactic SN Ia rate of $0.007$ 
    yr$^{-1}$,][section 4]{BM12}.
   This number is within observational constraints, and in fact our
   example system exhibits physical properties similar to those of the helium nova V445 Puppis \citep{Kat08}.}  
The mass transfer rate soon after enters the nuclear time-scale regime, 
and as it decreases helium burning continues to be stable on the CO~WD. 
Once the mass transfer rate drops
below $\dot{M}_{\rm tran} = 1.74 \times 10^{-6}$ M$_{\odot}$ yr$^{-1}$
(still $t=128$ Myr; WD mass $1.08$ M$_{\odot}$),
helium burning is no longer stable (Equation 4, Kato \&
Hachisu 2004) and thus not fully efficient.  
The primary continues to accrete
another $0.11$ M$_{\odot}$ (total mass transfer including
Eddington-limited phase lasts 0.15 Myr), though RLOF shuts off when the orbit has
expanded to $a=1.91$ R$_{\odot}$ (orbital period of 5.2 hr).  
At this point, $m_{\rm p}=1.19$ M$_{\odot}$ and $m_{\rm s}=0.78$ M$_{\odot}$,
while the secondary's helium core is almost completely burned to CO 
($m_{\rm core}=0.76$ M$_{\odot}$).  
Within $10^{5}$ yr (after losing ${\sim}0.001$ M$_{\odot}$ in stellar 
winds) the secondary becomes a CO~WD ({\bf stage
  IX}). The two WDs evolve toward smaller orbital separations over the
next Gyr under the influence of gravitational wave radiation, and
merge at $t=1259$ Myr ({\bf stage X}), with a combined mass
of $1.96$ M$_{\odot}$.  Such a delay time (1259 Myr) is typical of
double degenerate mergers -- they are predicted to produce more SNe~Ia at
early times, peaking at a few hundred Myr though still producing a significant
number of events ${\gtrsim}1-3$ Gyr \citep[e.g. fig. 1,][]{RBF09}. 

The system described above is typical of an evolutionary channel in which
the critical accretion phase onto a WD (hereafter stage VIII,
cf. Fig. 2) occurs.  
For the evolutionary model under consideration \citep[$\alpha_{\rm CE}\lambda = 1$ of ][]{Rui11},
we find that 43\% of {\em all} CO+CO~WD merger progenitors undergo such 
an accretion phase
during which the primary WD (e.g. the first-formed and {\em more massive}
WD) accretes an additional $0.001-0.45$
M$_{\odot}$ after its formation (in the majority of cases it gains 
${\sim}0.2 \pm 0.1 $ M$_{\odot}$).   
In addition to these 43\%, 9\% of merger progenitors encounter the opposite situation: 
the secondary WD (e.g. the first-formed but {\em less-massive} WD) 
undergoes helium accretion.\footnote{These secondaries only gain a small amount of mass 
and they never become the more massive (exploding) WD.}
The remaining 48\% of
progenitors do not encounter a stable mass accretion phase while one of the
stars is a WD.  

Different physical properties of a binary at birth (masses, mass ratio
and separation/eccentricity) influence the ensuing evolution.  We find
that for the (43\%) group of primary interest, typical ZAMS masses are 
somewhat large, peaking at ${\sim}4-5$ M$_{\odot}$ (though the entire range is
$2-7$ M$_{\odot}$), and initial mass ratios 
$q_{0}$(= more massive star / less massive star) range 
${\sim}1-2$ with a peak at ${\sim}1$. 
For cases where the WD-accretion phase occurs 
on the secondary WD (9\%), ZAMS masses
are in the range ${\sim}2-3.5$ M$_{\odot}$ and thus result in
lower-mass WDs (in addition, $q_{0}$ for these systems
peaks at ${\sim}1.4$).  
For merger progenitors which do not undergo the WD-accretion
phase (48\%), ZAMS masses are mostly ${\sim}2-3$
M$_{\odot}$, and $q_{0}$ strongly peaks at unity.  
Hence, a large fraction of these stars evolve at roughly equal rates 
and in many cases, they reach the short-lived helium star phase 
at roughly equal times, so stage VIII does not occur.

\section{Condition for prompt detonation}
\label{sec:qc}

Binary population synthesis calculations 
provide a predicted distribution of physical parameters for
merging CO~WD pairs. However, the prompt detonation explosion
mechanism of \cite{pakmor10,pakmor11,pakmor12} is not expected to be
realised for all CO~WD mergers.
As discussed by \cite{pakmor10,pakmor11}, 
conditions favouring a prompt detonation are most plausible
when the ratio of the secondary WD mass ($m_{\rm s}$) to the primary
WD mass ($m_{\rm p}$) is high (approaching $1$).
In a study of mergers with $m_{\rm p} = 0.9$~M$_{\odot}$,
\cite{pakmor11} determined that the critical 
mass ratio ($q_{\rm c} = m_{\rm s} / m_{\rm p}$)
above which explosion is
plausible is $q_{\rm c}{\sim}0.8$. The critical $q$-value, however,
may depend on the mass of the primary, likely having a value that
decreases with increasing $m_{\rm p}$ 
\citep[for higher-mass primaries, the densities are higher and the merger
is more violent making it easier to reach the conditions needed for
triggering a detonation,][]{Sei09}.
Presently, sufficiently detailed simulations to determine how $q_{\rm
  c}$ will depend on primary mass have not been carried out.
Therefore, in this
study, we will adopt a simple parametrization of $q_{\rm c}$:
\begin{equation}
q_{\rm c} = \rm{min} (0.8 \left( {\frac{m_{\rm p}}{0.9 {\rm
        M}_{\odot}}} \right)^{-\eta}, 1.0)
\end{equation}
where $\eta$ is a parameter 
that we will use to explore how $q_{\rm  c}$ influences our 
findings.

\section{Primary mass/peak brightness relation}
\label{sec:bright}

Ideally, synthetic observables for a suite of three-dimensional WD merger
models would be used to determine the relationship between the system
parameters and the explosion brightness. Unfortunately, such simulations
are too computationally expensive for this to be practical in this study.
However, as argued in Section~\ref{sec:intro},
the peak brightness of a prompt detonation merger model
is directly related to the mass of the more massive WD ($m_{\rm p}$); in
the merger, the $^{56}$Ni is mostly synthesized in the detonation of the
primary.
We therefore estimate the brightness of a merger from a simple
one-dimensional model for a detonation of a WD with mass equal to that of
the primary.

To quantify the relationship between the primary mass and
the peak bolometric magnitude ($M_{\rm bol}$), we 
carried out a series of simulations for detonations of single CO~WDs
with a range of WD masses ($m_{\rm WD}=m_{\rm p}$). 
These simulations were computed in exactly the same manner as those
described by \cite{Sim10}. 
We constructed hydrostatic models
for CO~WDs 
and simulated detonations of these WDs using our
SN~Ia explosion code. In all cases, the detonations were centrally
ignited and their propagation followed using the level-set technique 
\citep{GN05,Fin10}. Nucleosynthesis in the explosions was computed
in a post-processing step 
using our standard tracer particle approach \citep{travaglio04,Sei10}
assuming an initial $^{12}$C/$^{16}$O/$^{22}$Ne composition of
47.5/50/2.5 \% by mass (roughly appropriate for solar
metallicity). For each explosion model we 
performed radiative transfer simulations with ARTIS
\citep{sim07,kromer09} to predict synthetic light-curves (adopting the 
cd23\_gf-5 atomic dataset of \citealt{kromer09}).

\begin{table}
\begin{center}
\caption{The peak bolometric magnitude derived from our 1D sub-Chandrasekhar mass pure detonation models for CO~WDs for a range of WD masses ($m_ {\rm WD}$).}
\begin{tabular}{ll} \hline
$m_{\rm WD}$ [M$_{\odot}$] & $M_{\rm bol}$ [mag] \\ \hline
0.81 & -13.75\\
0.88 & -16.74\\
0.97 & -18.20\\
1.06 & -18.85\\
1.10 & -19.03\\
1.15 & -19.22\\
1.28 & -19.49\\ \hline
\end{tabular}
\end{center}
\end{table}

\begin{figure}
\epsfig{file=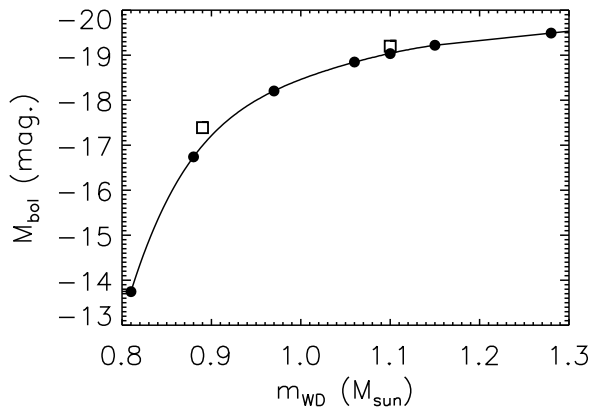,width=8.8cm}\\
\caption{The $m_{\rm WD}$-$M_{\rm bol}$ relationship derived from the pure
detonation models in Table~1. The filled circles indicate the data derived
from the models and the solid line shows the polynomial interpolation we
adopt in our analysis. The squares indicate the results from the full
merger simulations described by Pakmor et al. (2010; subluminous) and
Pakmor et al. (2012; $M_{\rm bol}{\sim}-19$).}
\label{fig:fit}
\end{figure}

Our chosen values of $m_{\rm WD}$ and the derived values of $M_{\rm bol}$ are given in Table~1.
Fig.~\ref{fig:fit} shows the $m_{\rm WD}$-$M_{\rm bol}$ relationship
derived from the models.  
As expected, the models define a smooth,
monotonic relationship between brightness and mass (see e.g. \citealt{Sim10}). 
For the analysis below, we have made a polynomial interpolation
between the data points defined by the models to obtain a simple
functional form for the $m_{\rm WD}$-$M_{\rm bol}$ relationship. This
interpolated relationship is also shown in Fig.~\ref{fig:fit}. 
(see Appendix B for coefficients of the fit).

Recently, \cite{pakmor12} presented results from a complete simulation
of a prompt-detonation model for the merger of a CO~WD pair with
$m_{\rm p} = 1.1$ and $m_{\rm s} = 0.9$~M$_{\odot}$.
The angle-averaged peak brightness for this model is
indicated in Fig.~\ref{fig:fit}. As expected, its
peak brightness is very close ($\Delta M_{\rm bol} \sim 0.2$ mag) 
to that of our equivalent single WD detonation model.
We also show the angle-averaged peak $M_{\rm bol}$ for the complete
merger simulation of \cite{pakmor10}. Again, the agreement with our relation
is good, although the difference is slightly larger
($\Delta M_{\rm bol} \sim 0.4$ mag).
These comparisons lend credence
to the approach we adopt to estimate the peak brightness of WD merger models
and suggest that our method is accurate to a level of several tenths of a
magnitude.

\section{Results}
\label{sec:results}

\subsection{Distribution of brightness}

To convert the population synthesis data to a distribution of peak
brightness, we first select all systems from the binary population
synthesis calculation that lead to CO+CO mergers. 
From these we identify those systems that will explode based
on our mass ratio cut ($q > q_{c}$, and chosen values of $\eta$; 
Section~\ref{sec:qc}). Finally, we apply our 
$m_{\rm WD}$-$M_{\rm bol}$ relationship (Section~\ref{sec:bright}) to
derive the peak brightness for each system from its primary WD mass.
We exclude from our
analysis systems in which the primary WD is $m_{\rm p} <
0.8$~M$_{\odot}$, for two reasons. First,
for such small primary masses the low densities make it 
extremely challenging to ignite the detonation.
Second, at low mass the $^{56}$Ni yield is expected to be too small to
give rise to a bright transient even if a prompt detonation were to
occur (see Table~1). 
This
effectively removes the majority of systems associated with the
low-$m_{\rm WD}$ peak (${\sim}0.65$ M$_{\odot}$) from our
analysis: those systems are not promising candidates for the prompt
detonation scenario (although it has been suggested that explosion of such systems
might still occur, \citealt{vanKerkwijk10}; but see also \citealt{She12}).

\begin{figure}
\epsfig{file=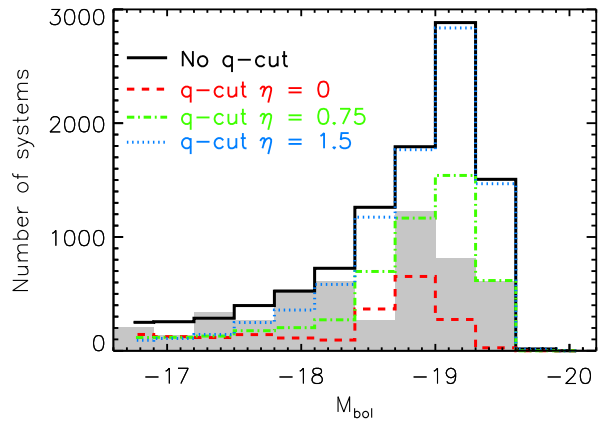,width=8.8cm}\\
\caption{The predicted distribution of peak bolometric magnitude ($M_{\rm
bol}$) obtained by combining the primary WD masses from the model
of \citet{Rui11} with our $m_{\rm WD}$-$M_{\rm bol}$
relationship (Fig.~\ref{fig:fit}).
The black line shows the results assuming that all mergers
successfully lead to a prompt detonation.  
The coloured lines show the effect of
imposing cuts on the $q$-value of the merger (requiring that $q > q_{\rm
c}$) for different values of $\eta$ (see text).
The grey histogram shows the absolute magnitude distribution derived from
a volume-limited sample of SNe~Ia by \citet[][see their fig.~5]{li11}. To facilitate
comparison of the shapes, the observational data have been 
scaled to match the number of systems in our calculation for $\eta = 0.75$.}
\label{fig:mag-dist-std}
\end{figure}

Our computed brightness distribution is plotted in Fig.~\ref{fig:mag-dist-std}
for three representative values of the $\eta$-parameter in our
prescription for $q_{c}$ (see Equation~1).
We have restricted $\eta$ to be positive, ensuring that 
the critical $q$-value decreases with increasing $m_{\rm p}$ (except
for the case where $\eta=0$, where $q_{\rm c}=0.8$ for all values of
$m_{\rm p}$).  We find that for $\eta {\ge}1.5$, nearly all of the primaries 
at the high-mass end are included, so adopting larger values of $\eta$
is unnecessary.  Thus, we choose two values of $\eta$ which bracket the 
physically-realistic extremes: $\eta=0$ and $\eta=1.5$, and a third
value $\eta=0.75$ (the midpoint), simply for illustrative 
purposes.\footnote{We note that for our evolutionary example of a violent
  merger formation channel described in Section 2.3 
  ($q_{\rm merger} = m_{\rm s}/m_{\rm p}=0.774/1.185=0.653$), 
the necessary criteria leading to a SN~Ia are satisfied for 
$\eta > 0.74$: 
$\eta = 1.5 \rightarrow q_{\rm c}=0.530$;  
$\eta = 0.75 \rightarrow q_{\rm c}=0.651$.}
  We also show the result without any $q$-cuts (i.e. all population
synthesis data).

As shown in Fig. 5,
our $m_{\rm WD}$-$M_{\rm bol}$ relationship  
converts the higher mass peak of the $m_{\rm p}$ distribution
(Fig.~\ref{fig:mass-dist-std}, solid line) into a
brightness distribution with a clear peak 
around $M_{\rm bol} {\simeq}-19$~mag, a short tail of brighter events, 
and a long tail stretching to fainter magnitudes. The
amplitude of the peak and to some degree its location depend on our choice of $\eta$.
For $\eta = 1.5$ (or higher), the
high $M_{\rm bol}$ part of the distribution is almost insensitive to
the choice of $\eta$ but for smaller values (e.g. $\eta = 0$), a
significant fraction of the high $m_{\rm p}$ systems are excluded.
However, $\eta$ does not qualitatively affect our findings: the peak 
around $M_{\rm bol} {\sim}-19$~mag is present in all cases -- 
it is a consequence of the underlying shape of the 
$m_{\rm p}$ distribution in merging systems and the
steepness of our $m_{\rm WD}$-$M_{\rm bol}$ relationship.

\subsection{Comparison to observations}
\label{sec:obs_comp}

Recently, \cite{li11} presented pseudo-observed luminosity functions
for SNe~Ia based on a sample of objects within 80 Mpc from the Lick
Observatory Supernova Search \citep[LOSS, see
e.g.][for details]{leaman11}. We compare the shape of their
volume-limited luminosity function to the results of our calculations
in Fig.~\ref{fig:mag-dist-std}. In this comparison, we include all
74 of their SNe~Ia (regardless of sub-type and host galaxy type) and,
for convenience, have arbitrarily rescaled the distribution to match the total
number of SNe~Ia to that of our calculation for $\eta = 0.75$. 

This comparison suggests that the prompt detonation violent merger
scenario could give rise to a brightness distribution of SNe~Ia that
is fairly compatible with observations.  In particular, the
standard model of \cite{Rui11} predicts a large fraction of
explosions within a fairly narrow
range of peak magnitude ($\Delta M_{\rm bol} \sim 1$~mag) centred
around $M_{\rm bol} {\sim}-19$~mag, as is observed.

There are a number of important caveats to consider when interpreting
the comparison in Fig.~\ref{fig:mag-dist-std}. First, the supernovae in
the \cite{li11} luminosity function have been corrected for Galactic
extinction but not for host galaxy reddening, thus variations in extinction
from object to object will be responsible for some component of the
tail to lower peak magnitudes in the observed sample. Second, the
\cite{li11} data are not derived from complete sets of colour
light-curves but are available only for the $R$-band. However, this is
not expected to significantly affect our results around the peak of
our distribution: in the complete
merger simulation presented by \cite{pakmor12}, the difference in peak
bolometric and $R$-band magnitudes was $M_{\rm bol} -
M_{\rm R} = 0$. We note that $M_{\rm bol} -
M_{\rm R}$ is expected to become positive for sub-luminous mergers 
but for the systems of interest, this should not be large effect 
compared to the bin size in Fig.~\ref{fig:mag-dist-std}
(in the simulation of \citealt{pakmor10}, $M_{\rm bol}-M_{\rm R} {\sim}0.4$ mag.).
Third, our analysis neglects any influence of observer
orientation on the explosion brightness -- the \cite{pakmor10,pakmor12}
simulations suggest that orientation may cause variations of around
$\pm 0.4$~mag.
Nevertheless, although quantitatively important, none of the effects
is expected to qualitatively change our result.

\subsubsection{Brightness and delay time}
\label{subsec:Mp-DTD}

\begin{figure}
\epsfig{file=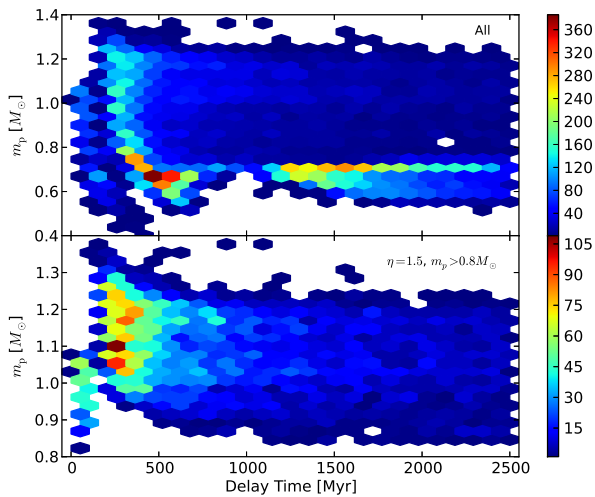, width=9.8cm }\\
\caption{Primary WD mass ($m_{\rm p}$) vs. delay time up to 2500 
Myr for {\sc StarTrack} CO+CO~WD mergers. The upper panel shows all 
systems while the lower panel only includes systems that pass the 
$q_{\rm c}$ criterion with $\eta = 1.5$ and $m_{\rm p}>0.8 M_{\odot}$.
% Relative numbers are indicated in colour. 
Note the different scaling for the two plots. 
It is evident that the most massive
 primaries preferentially merge
  at prompt delay times: for both panels $\gtrsim${95\%} of systems with $m_{\rm
    p}>1.3\,M_{\odot}$ have delay times $<1$ Gyr.  Mergers with $m_{\rm
    p}>1.3\,M_{\odot}$ make up 0.2\% of all CO+CO mergers and 0.5\% of $\eta =
  1.5$, $m_{\rm p}>0.8 M_{\odot}$ mergers.
We also note that the first progenitors to
  explode are not from binaries with the most massive primaries (see
  text for details).  
}
\label{fig:WDmass-dtd-dens}
\end{figure}

We note the important finding that, while fairly massive WDs with masses of
${\sim}1.2$ M$_{\odot}$ may explode at any delay time ${>}150$ Myr, the most massive ones (${\ge}1.3$ M$_{\odot}$) only explode
as prompt events, with rare exceptions 
(see Figure~\ref{fig:WDmass-dtd-dens}).  This finding supports the observational evidence for more
luminous SNe Ia occurring among younger stellar populations \citep{Bra11}.
At the same time however, the events with the shortest delay times ($<100$ Myr) 
are not from the most massive primaries: these occur via rarer formation channels where
(usually) the secondary initiates two CE events.  These `ultra-prompt'
double CE systems do not undergo phase VIII mass accretion, though they are
still massive enough to be considered SNe Ia progenitors in our
model.  The difference in evolutionary
timescale between these double CE events and all of the other formation channels is the cause of
the dearth of mergers at ${\sim}150$ Myr in Figure~\ref{fig:WDmass-dtd-dens}.  For the
majority of the events at later delay times only one CE is
encountered, as is expected \citep{RBF09,Men10}.

\begin{figure}
\epsfig{file=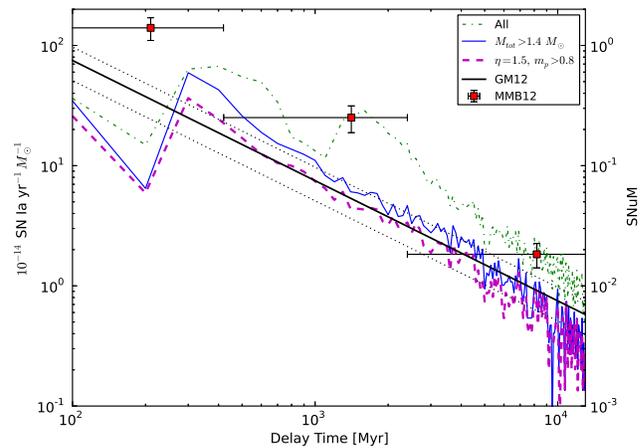, width=8.8cm }\\
\caption{{\sc StarTrack} smoothed DTD of all CO+CO mergers
  (green dots), CO+CO mergers where $M_{\rm tot}>1.4 M_{\odot}$
  (blue solid), and CO+CO~WD mergers which satisfy our criteria
  $\eta=1.5$ and $m_{\rm p}>0.8 M_{\odot}$ (magenta dashes).  The small
  fluctuations are due to Monte Carlo noise.  Over-plotted with red
  squares is the best-fit DTD derived from SDSS data, the three time bins
  are indicated by horizontal error bars: $t<0.42$ Gyr, $0.42<t<2.4$
  Gyr and $t>2.4$ Gyr \citep[see][MMB12]{MMB12}.  The smooth black solid line
  shows the best-fit DTD from \citet[][GM12]{GM12}, with systematic
  uncertainties (thinner dotted lines).  The right-axis shows units in SNuM: number of SN Ia
per $10^{10} M_{\odot}$ formed in stars per century.}
 \label{fig:dtds-log}
\end{figure}

\subsubsection{Rates and DTD}
\label{subsec:rates}

We have not attempted to calibrate the population
synthesis event frequency of CO+CO violent mergers to compare directly
with the SN~Ia rates of \citet{li11}. 
To do so would require us to adopt a model for the star-formation history
associated with their volume-limited sample, which would 
introduce unwanted uncertainties.
Nonetheless, we can compare our theoretical DTD to that
  from observations.  
Merging CO+CO~WD pairs are currently 
the favoured theoretical scenario for producing 
a relatively high number of SN Ia progenitors compared to other
long-standing progenitor models \citep{IT84,Yun94,RBF09}. 
In Figure~\ref{fig:dtds-log} we show three theoretical DTDs
for WD mergers assuming a binary 
fraction\footnote{For stars of spectral type B3-B7, which are typical
  progenitors of violent WD mergers, the binary fraction is estimated
  to lie
  between 40-100 \% \citep{HNC11}.  We do not consider higher
  multiplicity here.} of 70\%: 
All CO+CO~WDs, mergers for which $M_{\rm tot}>1.4$ $M_{\odot}$ (DDS), 
and systems which meet our criteria $\eta = 1.5$ and 
$m_{\rm p}>0.8$ $M_{\odot}$ (violent mergers). 
Additionally we show two of the most up-to-date
observationally-recovered DTDs from the literature: \citet[][SDSS II
galaxies; fig. 1]{MMB12} and \citet[][field rates; fig. 12]{GM12}.

The DTD from violent mergers (magenta dashed line) 
compares rather well with the most recent observational DTD
estimates.  At long delay times ($t>2.4$ Gyr)
the violent merger DTD falls within 
the observational uncertainties of \citet{MMB12}. 
Also, the model agrees very well 
with the best-fit DTD from 
\citet{GM12} for delay times above 300 Myr, 
implying that violent mergers might be a 
dominant contributor to SNe~Ia in field galaxies   
\citep[see][for a comparison of observational DTDs from
recent works]{GM12}.  

Up until now, there has been a 
notable discrepancy between observed
rates of SNe Ia -- in particular at early delay times -- and the rates 
calculated from population synthesis.  Typically, binary population
synthesis rates (as a function of delay time) were too low to explain
the observed numbers.  This discrepancy remains an 
open issue, though some possible reasons have been
discussed\footnote{Observational samples which
    include massive elliptical galaxies tend to exhibit DTDs which
    have higher SN rates
    than volume-limited samples, D. Maoz, private communication 2012;
    see also \citet{MMB12}.} 
\citep[see][sect. 4.1; see also Mennekens 2010 who point out that
rotation may increase the likelihood for explosion in WD
mergers]{NTB12,Rui11}.  
However, the new observationally-derived DTD of \citet{GM12} has 
a lower amplitude (implying lower SN~Ia rates) than those from 
previous observations.  This brings our theoretical DTD, as 
well as some theoretical DTDs from other groups, 
into better agreement with observations.  
It should be noted that \citet[][sect. 6]{GM12}  
mention that their best-fit power-law
DTD might be underestimated, since it was derived using
predominantly old galaxies.  

%However, the agreement between the observations and the
%theoretical predictions shown here implies that previous calculations of 
%double degenerate merger rates from other groups are now also brought
%into better agreement with the observations.  

\section{Discussion}
\label{sec:discuss}

We have found that, in the standard model of \citet{Rui11}, a certain 
formation channel is critical for creating massive primary WDs.  
In Section~\ref{subsec:binev} we described the typical evolution of a binary which
produces two merging CO~WDs where the primary WD is particularly
massive due to an important evolutionary phase: it accretes
(in that case $0.35$ M$_{\odot}$) from a helium-rich companion star. 
Within the model framework of \citet{Rui11}, such a phase of RLOF
occurs for a significant fraction (43\%) of the CO+CO~WD
merger progenitors.

We have mapped the (exploding) primary WD 
masses for all CO+CO~WD mergers on to magnitude-space using 
our $m_{\rm WD}$-$M_{\rm bol}$ relation, 
and have presented the results for a series of $q$-cuts 
(Fig.~\ref{fig:mag-dist-std}). 
Regardless of the $q$-cut used, the shape of our theoretical 
$M_{\rm bol}$ distribution for merging
primary WDs is quite striking, since it matches well with 
the observed $M_{\rm bol}$ distribution for SNe~Ia of \citet{li11}.
{\em Our results indicate that double degenerate mergers exploding in the
violent merger scenario show exceptional
promise as progenitors of SNe~Ia.} 

However, despite these very encouraging results, we point out in the
next two sub-sections some 
important issues to keep in mind, which could have a significant
effect on the evolutionary outcomes for close binaries, in turn
affecting the impact of these new findings.

\subsection{A key assumption: Common Envelope parametrization}

Different assumed CE parametrizations have a 
critical impact on which evolutionary channels lead to the formation 
of close binaries, as well as how often they form and how often they 
merge \citep{Dom12}. So, we now discuss how the results found above are sensitive to 
the choice of the adopted CE formalism and CE efficiency.  
 
Considering a lower CE efficiency such as $\alpha_{\rm CE}=0.5$
with  $\lambda=1 \rightarrow \alpha_{\rm CE} \lambda=0.5$, which is often used in 
population synthesis, we find that the 
stage VIII accretion phase is indeed 
encountered though not as often as in the standard model (33\%
vs. 43\%). 
Quantitatively speaking, the overall merger rates from population
synthesis calculations for this CE model are about a factor of two 
lower than the standard model \citep{RBF09}, 
since a larger number of progenitors which make CO+CO~WD mergers 
in the standard model will merge before a double WD is formed 
in a lower CE efficiency model. 
Qualitatively however, there is little difference between
the two models, and in the $\alpha_{\rm CE} \lambda=0.5$ model we 
still expect such a `stage VIII' accretion phase to occur leading 
(favourably) to more massive primary WDs.   

For the case of very low CE efficiency, such as that investigated
in \citet{Rui11} ($\alpha_{\rm CE}=0.25$ with $\lambda=0.5 \rightarrow$
$\alpha_{\rm CE}\lambda=0.125$), stage VIII 
is almost never encountered: less than 1\% of binaries go
through this evolutionary channel ($>99$\% of primary WDs do not
experience accretion before the merger).  In that calculation, the 
formation channels that do lead
to CO+CO~WD mergers are altogether different from our standard case.    
For the $\alpha_{\rm CE} \lambda=0.125$ model, binaries which end up merging as CO+CO~WDs
generally start out with larger semi-latera recta 
($a \times (1.0-e^{2})$) of ${\sim}50-250$ R$_{\odot}$, 
while for the standard model initial semi-latera 
recta typically range from $20-100$ R$_{\odot}$.
For these merging WD pairs, the primary WDs are in general less massive
than their standard model counterparts.

The 0.2\% of CO+CO~WD merger progenitors which do encounter 
the stage VIII accretion phase in the $\alpha_{\rm CE}\lambda=0.125$
model are born with near-equal masses, 
are closer at birth (semi-latera recta $20-100$
$R_{\odot}$), and the first-formed WD always forms either during or
after the CE (in contrast to our standard example, where it forms
prior to the CE event).  
In all of these systems, there is a mass ratio reversal during the 
evolution, which is  
significant enough to allow for 
rapid rejuvenation of the secondary 
and the secondary is the first star to 
evolve into a WD (this process is also encountered on occasion 
in the standard model; Section~\ref{subsec:binev}).  
However, these binaries do not meet our criteria 
for SN~Ia progenitors: the most massive primary WD (at merger) 
in the 0.2\% group is $0.7$ M$_{\odot}$.  This stems from
the fact that the ZAMS progenitor masses for this
channel are quite low ($\sim 1.9$ M$_{\odot}$).  
In any case, as demonstrated in \citet{Rui11} the expected rate of
merging CO+CO~WDs is extremely low for this CE model.    

If we consider the `$\gamma-$formalism' for CE evolution 
\citep[e.g.][]{Nel00}, which was assumed for all CE events 
for model G1.5 in \citet{Rui11}, stage VIII is also not
encountered very often.  
The typical formation channels which lead to double WDs from the
standard model do not make double WDs in the `$\gamma-$formalism'
model, since for the latter the stars have larger orbits
post-CE, and will not merge within a Hubble time.
Four \% of primary CO~WDs
undergo stage VIII mass accretion, while for 3\% it is the secondary
WD for which the phase occurs.  Both groups have $q_{0}={\gtrsim} 1.2$, 
and often the first interacting event is a CE,
followed by a second CE after the primary has evolved into a CO~WD.  
Both CE events enable the stars to be on a close enough orbit such
that RLOF between a slightly evolved helium star and the primary WD
is possible.  In terms of frequency of events, the corresponding DDS model
of \citet{Rui11} yields a rate which is 10\% of that of the standard
model, thus we would expect a smaller fraction of potential 
SN~Ia progenitors from violent mergers to be produced via the G1.5
model. 

Not surprisingly, binary population synthesis predictions for double
degenerate mergers are rather sensitive to the assumed CE parametrization.
In summary, a small reduction of CE efficiency ($\alpha_{\rm CE}\lambda = 1 \rightarrow 0.5$)
does not qualitatively affect our results. However a very low efficiency 
($\alpha_{\rm CE}\lambda = 0.125$) or use of the $\gamma$-prescription
alone dramatically
reduces both the total merger rate and the fraction of systems that undergo
the stage VIII accretion phase.
We note additionally that using a higher CE-ejection efficiency 
\citep[$\alpha_{\rm  CE}>1$,][]{MY12} should lead to a larger number of 
WD mergers.

\subsection{WD mass growth: implications}

\begin{figure}
\epsfig{file=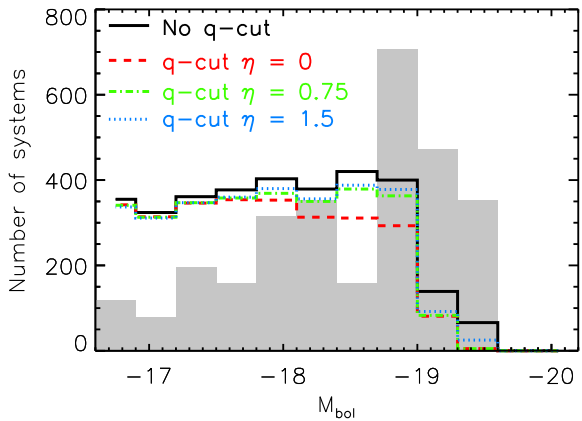,width=8.8cm}\\
\caption{As Fig.~\ref{fig:mag-dist-std} but using the WD masses as
  they were {\it at birth}.  No matter which $q$-cut is used, no
  theoretical $M_{\rm bol}$ distributions exhibit a shape similar
  to the absolute magnitude distribution of SNe~Ia from \citet{li11}.}
\label{fig:mag-dist-birth}
\end{figure}

As an experiment to confirm that
the mass-growth of the primary WD after it 
is born is key to our findings, in 
Fig.~\ref{fig:mag-dist-birth} we show results equivalent to 
Figure~\ref{fig:mag-dist-std} 
but using the {\sc StarTrack} WD primary masses 
{\it at their birth} 
(dashed line in Figure~\ref{fig:mass-dist-std}) 
rather than at explosion. 
It is immediately obvious that the shape of the observed 
absolute magnitude distribution of \citet{li11} is not reproduced in
this case.  Rather, for all $q$-cuts used, the theoretical $M_{\rm
  bol}$ distribution is rather flat from $-17$ to $-19$~mag, with a steep
drop-off at $M_{\rm bol}=-19$~mag.

While Fig.~\ref{fig:mag-dist-birth} is only an experiment and not 
a self-consistent calculation for our {\sc StarTrack} simulations, it 
may be 
roughly representative of predictions from alternative models.  
In particular, we note that other population synthesis codes predict a
{\em total} CO+CO~WD merger mass distribution that is fairly smooth 
\citep[Nelemans~2010\footnote{{\em Workshop on ``Observational Signatures of Type Ia
Supernova Progenitors''}, Leiden 2010 (http://www.lorentzcenter.nl/lc/web/).}; see also][fig. 1]{BT10},
in contrast to the total mass distribution from {\sc StarTrack} which has
a second peak as a consequence of stage VIII ({\rm cf}. Fryer et al.~2010 fig. 3).
We have identified an evolutionary channel that produces fairly
massive primary WDs via helium-accretion, though it remains to be 
established whether such a prediction is confirmed 
by future observations (see Footnote 3).
In addition, improved input physics 
from models of interacting binaries 
that incorporate detailed stellar structure -- in particular for
helium star models -- are needed to properly address the issue of 
mass transfer/accretion in close binaries \citep[see e.g.][]{Egg11}.  

\subsection{Ultra-prompt mergers}

 As mentioned in section~\ref{subsec:Mp-DTD}, the SNe Ia from WD mergers with delay times $<
 100$ Myr (`ultra-prompt') are formed by way of a double CE: the
 initially less massive star loses its H-rich envelope while in the
 Hertzsprung gap or on the giant branch, and then later loses its
 He-rich envelope when it is a somewhat-evolved helium star.
 We find that 0.5\% of all CO+CO mergers which have primary WD
 masses $> 0.9 M_{\odot}$ merge less than 1000 yr after the last CE.
 The fraction is raised to 4\% if the merger occurs within 10,000 yr
 after the last CE.  This has very interesting implications for at least a couple of
 reasons.  The post-explosion spectra of these mergers are likely to
 show signatures of circumstellar material -- thus far thought to be a
 likely signature of a single degenerate scenario SN Ia
 \citep{Ste11}.  On the other hand, since the primary WD will still be quite hot
 upon merging, these binaries may meet the necessary criteria for
 producing `core degenerate' mergers \citep[see][]{IS12}, which might
 produce SNe Ia -- or some other bright transient -- at a delay time
 larger than the merger time shown in Fig.~\ref{fig:dtds-log}.  In any case,  
these types of mergers might contribute a small fraction to the 
population of sub-luminous SNe Ia.

\subsection{Violent vs. non-violent mergers}

\citet{Che12} used population synthesis 
to calculate rates of SNe Ia from CO+CO~WD mergers 
and imposed some additional constraints,  
namely that in a dynamical merger some fraction of the
total mass would be lost from the binary, in turn lowering the
probability for a thermonuclear explosion.  
They found a possible upper limit of SNe coming from 
``violent mergers" to be about a factor of 5 lower than previous
estimates from population synthesis.  We would like to point out that
by our definition, the mergers considered in \citet{Che12} are not violent
mergers.  In the violent mergers of \citet[][]{pakmor12} -- on
which the \citet{Che12} dynamically-motivated restrictions are based -- 
no mass is lost in the merger itself.  In the violent merger, the 
mass transfer phase is very short-lived, and the explosion takes place
at the time of the merger at which point all of the matter is swept up
\citep[][]{pakmor12}. 
By contrast, the mergers assumed to undergo mass-loss 
considered by \citet{Che12}
are likely to exhibit properties of non-violent 
mergers, where a hot envelope is accreted onto the primary WD.
In such a case, a non-negligible amount of material might be ejected
from the binary (e.g. \citealp{Fry10}, but see also \citealp{Dan12}), 
though as already discussed, it is not clear that such systems will 
produce SNe Ia.

\section{Summary}
\label{sec:conclusions}

We find that the distribution of merging CO~WD pairs predicted by 
the {\sc StarTrack} binary evolution population synthesis code 
gives rise to a range of explosion brightnesses via the
prompt detonation (violent merger) mechanism \citep{pakmor10,pakmor12}
that is compatible with observed SNe~Ia (Fig.~\ref{fig:mag-dist-std}).
Further, the DTD (Fig.~\ref{fig:dtds-log}) agrees well with 
the observed DTD fit from \citet{GM12}, and as shown by 
Fig.~\ref{fig:WDmass-dtd-dens} (lower panel), 
even gives rise to brighter SNe Ia preferentially at
short delay times -- a trend that is corroborated by SN Ia 
observations \citep[e.g.][]{WH12}.

We have identified that the good agreement with the brightness
distribution of \citet{li11} depends critically on a particular 
evolutionary phase during 
which the first-formed WD accretes mass 
from a companion which 
is a slightly evolved naked helium star (stage VIII, Fig. 2).
In the standard population synthesis model of \cite{Rui11}, 
nearly half of the CO+CO~WD merger progenitors pass through 
such an accretion phase.

Alternatively, if the critical He-rich accretion phase 
is not readily realised in nature, then the predicted explosion
brightness distribution is not a good match to observations 
(Fig.~\ref{fig:mag-dist-birth}), likely indicating that  
a different explosion scenario must dominate and
drive the underlying shape of the 
SN~Ia observed brightness distribution.
Consequently, our results suggest that
detailed studies of helium accretion in binary systems will be a vital
step toward establishing the plausibility of the violent merger model for
a significant fraction of SNe~Ia.

\section*{Acknowledgments}

The authors thank the anonymous referee for comments, 
as well as Gijs Nelemans for helpful comments on 
this manuscript.  
AJR and SAS thank Richard Stancliffe and Brian Schmidt 
for stimulating discussions. 
AJR thanks Silvia Toonen, Joke Claeys and Nicki
Mennekens for insightful discussions 
on pertinent topics in binary star evolution and mass transfer, as
well as Dan Maoz and Rosanne di Stefano for informative conversation.
AJR also thanks Or Graur for kindly sending us data information.   
RP, MK and ST thank Brian Schmidt for hosting their visit to
{\it The Research School of Astronomy \& Astrophysics} of the 
{\it Australian National University}, during which part of this work
was carried out.    
The work of FKR and MF was supported by Deutsche 
Forschungsgemeinschaft via the Emmy Noether Program 
(RO 3676/1-1). FKR received funding from the ARCHES 
prize of the German Federal Ministry of Education and 
Research (BMBF).
IRS was supported by the graduate school ``Theoretical Astrophysics and 
Particle Physics'' at the University of W\"{u}rzburg (GRK 1147). 
SAS, MK, FKR and MF acknowledge financial support by the Group
of Eight/Deutscher Akademischer Austausch Dienst (Go8/DAAD)
Australian-German Joint Research Co-operation Scheme (`Supernova
explosions: comparing theory with observations.').
This research was undertaken with the support of resources provided by
the J\"{u}lich Supercomputing Centre in J\"{u}lich, Germany (project
HMU14) and the {\it NCI National Facility} in Canberra, Australia, 
which is supported by the Australian Commonwealth Government.

\appendix
\section{}

As mentioned in Section~\ref{sec:BPS}, we assume that when a
non-degenerate star accretes from its companion, 
$f_{a} = 0.5$, i.e. 
50\% of the mass lost from the companion is accreted while the 
other 50\% is lost from the binary 
\citep[see sect. 3.4 of][]{Bel08}.  
In nature, it is not clear how much mass may be lost/gained in such
a configuration, and the assumed value (and variability) of $f_{a}$ 
has an important
impact on the evolutionary outcome \citep[see the case study
of][]{Men10}.  More important than the fraction of mass
that is accreted might be in what manner angular momentum is lost
from the binary system during non-conservative mass transfer 
(see also Claeys et al., in preparation, for a comparison study 
of four population synthesis codes).  

For the current study, we used the published results from the standard
model of \citet{Rui11} for which $f_{a} = 0.5$ was assumed.  However, to obtain a general idea of how much our
results might change if a different $f_{\rm a}$ value had 
been used, we performed additional smaller-scale {\sc StarTrack} runs for $f_{\rm a}=0$
and $f_{\rm a}=1$.  We find that the relative frequency of CO+CO~WD 
mergers is highest for the fully non-conservative case: 
1.2 : 1 : 0.4 for a $f_{\rm a}$ fraction of $0$, $0.5$ and $1$,
respectively.  
For more massive WD mergers however, the trend changes.    
For CO+CO~WD mergers with $M_{\rm tot}>1.4 M_{\odot}$, 
the relative frequency is 0.8 : 1 : 0.3 and for mergers which meet the
criteria $\eta = 1.5, m_{\rm p}>0.8$ $M_{\odot}$, the relative frequency
is 0.5 : 1 : 0.3.  

Thus, assuming fully
conservative mass transfer onto non-degenerate accretors 
($f_{\rm  a}=1$) 
seems at face value to result in an overall decrease in the number of CO+CO
WD mergers.  Our test here shows that the overall rates for massive
CO+CO mergers might decrease at most by a factor of 3, though for
fully non-conservative mass transfer the overall CO+CO~WD merger rate
could increase by 20\%.  The trend found here is somewhat different
than that found by \citet{Men10}, however: one must be cautious and
accept that full evolutionary calculations (including other stages 
of binary evolution) are needed to
properly compute any relative change in our results, and such
a parameter study is beyond the scope of this paper.

\section{}

Here we give the necessary information to reconstruct the fit in
Figure~\ref{fig:fit}.  The 2-component fit was constructed with the
following equality: 
\begin{equation}
M_{\rm bol} = \sum\limits_{n=0}^{n_{\rm max}} a_{\rm n}\,m_{\rm WD}^{n} 
\end{equation}
where coefficients $a_{\rm n}$ for the two different mass regimes can 
be found in Table B1.  

\begin{table}
  \begin{center}
    \begin{minipage}[b]{0.65\linewidth}
      
      \caption{Coefficients for the polynomial and linear fits in
        Figure~\ref{fig:fit}.}
      \begin{tabular}{c}
        $0.8<m_{\rm WD}<1.15$ \\ \hline % & $1.15<m_{\rm WD}<1.4$ \\ \hline
        \begin{tabular}{ll}
          $n$ & $a_{\rm n}$\\ \hline
          0 & 2403.1822\\
          1 & -10943.254\\
          2 & 19821.934\\
          3 & -17966.776\\
          4 & 8140.1093\\
          5 &  -1473.6569\\ \hline
        \end{tabular}
      \end{tabular}
    \end{minipage}	
    
    \hspace{0.5cm}
    
    \begin{minipage}[b]{0.65\linewidth}
      \begin{tabular}{c}
        $1.15<m_{\rm WD}<1.4$ \\ \hline
        \begin{tabular}{ll}
          $n$ & $a_{\rm n}$\\ \hline
          0 & -16.827815\\
          1 & -2.0807692\\ \hline
        \end{tabular}
      \end{tabular}
    \end{minipage}
  \end{center}
\end{table}

    % \begin{tabular}{c} $\propto$2e (0-1)\\
%  \cite[]{Heggie75} \end{tabular} 

\section{}
Since on average the most luminous SNe Ia are often found among
relatively young stellar populations, while
less luminous SNe Ia are found among older stellar populations 
(Sect.~\ref{subsec:Mp-DTD}), we can 
explore how the shape of the violent WD merger peak brightness
distribution changes as a function of delay time.  

\begin{figure}
\epsfig{file=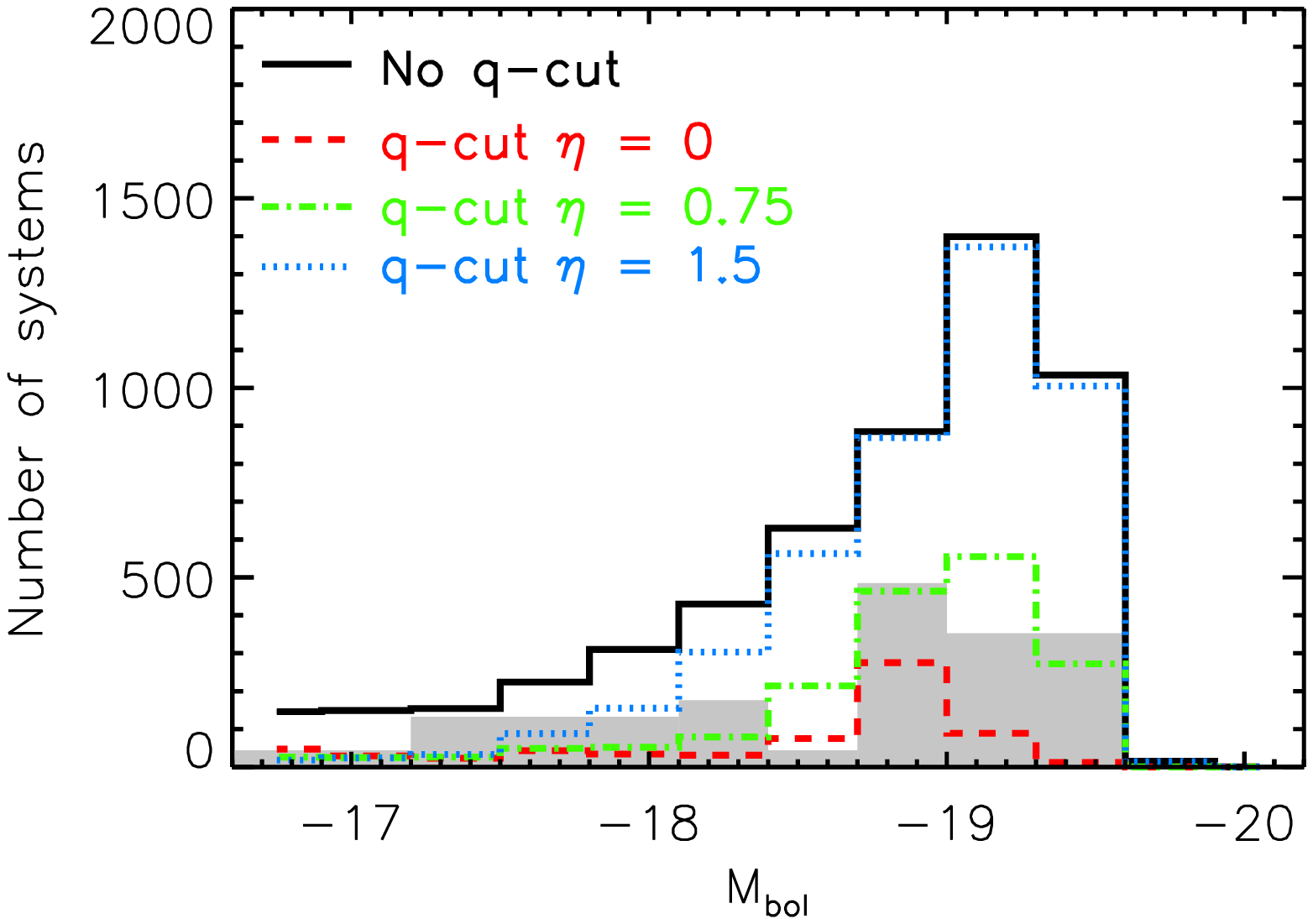,width=8.8cm}\\
\caption{Greyscale shows the peak brightness distribution of SNe Ia from Sb spiral and
  irregular galaxies as shown in \citet[][fig. 5]{li11}.
  Coloured lines show our theoretical peak brightness distributions
  for WD mergers with delay times less than 1
  Gyr.  Such a split
enables all `prompt' SNe Ia from violent mergers to be included in the
distribution, while still
including some SNe with intermediate delay times, which would be
expected in an observed sample involving spiral hosts (the 'Sb spiral \&
irregular' distribution of \citet{li11}).    
}
\label{fig:SbIrr}
\end{figure}

\begin{figure}
\epsfig{file=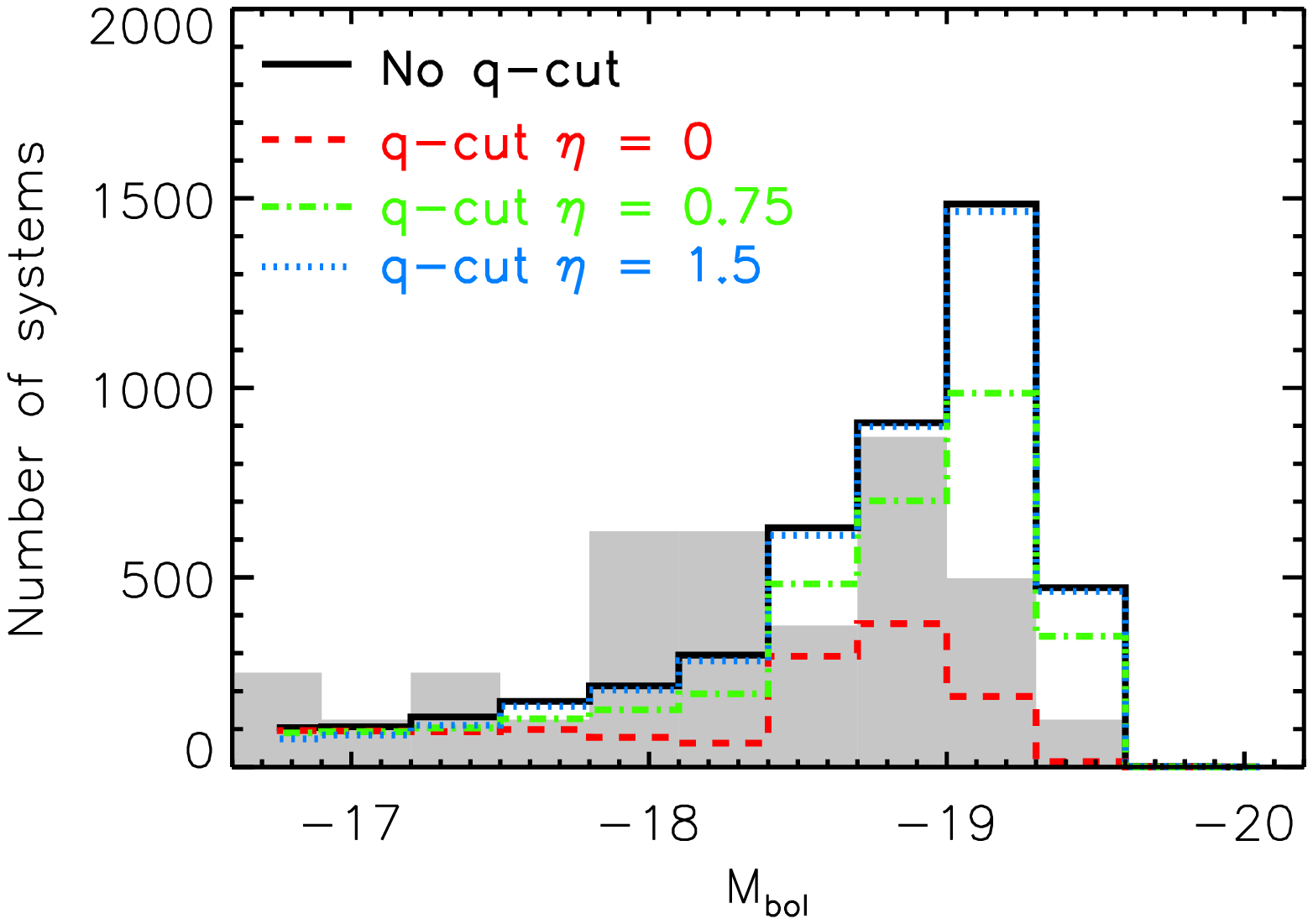,width=8.8cm}\\
\caption{Greyscale shows the peak brightness distribution of SNe Ia
  from elliptical and
  Sa spiral galaxies as shown in \citet[][fig. 5]{li11}. 
  Coloured lines show our theoretical peak brightness distributions
  for WD mergers with delay times greater than 1
  Gyr.  
}
\label{fig:ESa}
\end{figure}

We split our WD mergers (from Fig.~\ref{fig:mag-dist-std}) 
into two age groups: 
delay times less than $1$ Gyr and delay times greater than
$1$ Gyr.  The corresponding 
peak brightness distributions for these `younger' and `older' SNe Ia 
are shown in Fig.~\ref{fig:SbIrr} and Fig.~\ref{fig:ESa},
respectively. 
From the Figures, it is clear that the shapes of our theoretical brightness
distributions (coloured lines) do not change very much as a function of age.
Additionally, using a different time cut does not change the results
much either, since both bright and dim violent mergers occur at early 
and late delay times (see Sect.~\ref{sec:obs_comp}).  

We additionally compare our theoretical 
distributions to the two luminosity functions of \citet{li11} that are
separated by Hubble type.  
Li et al. (2011) split their luminosity functions into two groups
based on host galaxy morphology: SNe
Ia that are found in elliptical galaxies and spiral galaxies of type
Sa, and SNe Ia that are found in spiral galaxies of type Sb and
irregular galaxies (see their fig. 5).  
While delay times for
individual SNe are not available for the Li et al. (2011) sample, it
is reasonable to assume that the average delay times for
SNe Ia found among the group with elliptical galaxies will be larger
than the average delay times for SNe Ia found in the group which
contains irregular galaxies.   This is reflected in the slightly
higher proportion of low-luminosity events to high-luminosity
events in Fig.~\ref{fig:ESa}
compared to Fig.~\ref{fig:SbIrr} (greyscales).  
The \citet{li11} sample (both age groups)
contains SNe with different spectral properties, and we have included all 74
objects regardless of their spectral classification (normal,
high-velocity, 91T-like, 91bg-like, and 02cx-like).  

We note finally that the comparison of the theoretical brightness
distributions to the observational data 
is for illustrative purposes only, since 
an accurate comparison requires detailed information on the 
observed delay time distribution from the \citet{li11} sample.   
Furthermore, splitting the entire \citet{li11} sample as we have
done increases statistical noise: some bins only contain one SN 
event (see fig.~5 of Li et al.~(2011)).  
Despite these potential caveats,
the overall distribution shapes (observed and theoretical) show
similar trends, and compare relatively well.  In any case, 
it is possible (even likely) that other
formation channels and/or explosion mechanisms are also 
contributing at some level to the SN Ia
population at early and/or late delay times \citep[e.g.][]{kromer2012a}.

\label{lastpage}

\end{document}